\documentclass[
english, aps, prd,
superscriptaddress,
preprintnumbers,
nofootinbib,
twocolumn
]{revtex4-2}

\usepackage[T1]{fontenc}
\usepackage[utf8]{inputenc}

\usepackage{verbatim}
\usepackage{empheq}
\usepackage{amssymb}
\usepackage{amsmath}
\usepackage{amsfonts}
\usepackage{graphicx}
\usepackage{epstopdf}
\usepackage{siunitx}
\DeclareSIUnit\barn{b}

\usepackage{babel}
\usepackage{setspace}

\usepackage[colorlinks,linktocpage=true]{hyperref}
\hypersetup{citecolor=blue,linkcolor=blue,urlcolor=blue}

\newcommand{\be}{\begin{equation}}
\newcommand{\ee}{\end{equation}}

\newcommand{\Delprod}{\Delta_{\rm prod}}
\newcommand{\Deldec}{\Delta_{\rm dec}}
\newcommand{\Delexp}{\Delta_{\rm exp}}
\newcommand{\rmd}{\mathrm{d}}

\allowdisplaybreaks

\begin{document}

\preprint{TTP24-019,~P3H-24-040,~CERN-TH-2024-095,~TIF-UNIMI-2024-5}

\title{QCD corrections to Higgs boson production and \texorpdfstring{$H \to b \bar{b}$}{H -> bb~} decay in weak boson fusion}

\def\UR{Institute for Theoretical Physics,
    University of Regensburg, 93040 Regensburg, Germany}
\def\KIT{Institute for Theoretical Particle Physics,
	Karlsruhe Institute of Technology, 76128 Karlsruhe, Germany}
\def\MIL{ Tif Lab, Dipartimento di Fisica, Università di Milano and INFN,
	Sezione di Milano, Via Celoria 16, I-20133 Milano, Italy}
\def\CERN{Theoretical Physics Department, CERN, 1211 Geneva 23,
    Switzerland}

\author{Konstantin~Asteriadis}
\email{konstantin.asteriadis@ur.de}
\affiliation{\UR}

\author{Arnd~Behring}
\email{arnd.behring@cern.ch}
\affiliation{\CERN}

\author{Kirill~Melnikov}
\email{kirill.melnikov@kit.edu}
\affiliation{\KIT}

\author{Ivan~Novikov}
\email{ivan.novikov@kit.edu}
\affiliation{\KIT}

\author{Raoul~R\"ontsch }
\email{raoul.rontsch@unimi.it}
\affiliation{\MIL}

\begin{abstract}
\noindent
We study QCD corrections to the process where a Higgs boson is produced in weak boson fusion and then decays into a pair of massive $b$ quarks.
We find that typical experimental criteria used to identify $b$ jets in this process affect QCD corrections to the decay, making it necessary to account for them in the proper description of this process. Indeed, if corrections to the production and decay are combined, the fiducial cross section of the weak boson fusion process $p p \to H( \to b \bar{b}) jj$ is reduced by about $\qty{40}{\percent}$ relative to leading-order predictions, compared to just about $\qty{8}{\percent}$ if only corrections to the production process are considered.
We investigate the origin of these large corrections through next-to-next-to-leading-order and conclude that they appear because a number of independent moderately-large effects conspire to significantly reduce the fiducial cross section for this process.
\end{abstract}

\maketitle


\section{Introduction}
\label{sec:introduction}
Detailed studies of the Higgs boson will continue to play a central role in the physics program of the Large Hadron Collider (LHC) during Run 3 and the high-luminosity phase. They are expected to expand, refine, and consolidate the many important results in Higgs physics obtained since the Higgs boson discovery, which have given us a good understanding of the different properties of this particle and confirmed its central role in breaking electroweak symmetry.
With more data, improved detection capabilities, and novel analysis strategies, established Higgs signatures will be studied with greater precision, and rare signals may become accessible for the very first time.

Profiling the Higgs boson necessitates measurements of its couplings to matter and gauge fields, as well as its quantum numbers \cite{CMS:2022dwd,ATLAS:2022vkf}. Of the couplings, the Yukawa interactions are currently among the least known, and there is a huge interest in learning more about them. Indeed, currently, there is no information about the Yukawa couplings of fermions of the first generation, while those of the third generation fermions have been measured to about $\qty{20}{\percent}$ precision \cite{CMS:2022dwd,ATLAS:2022vkf}.
Because $H \to b \bar{b}$ is the dominant decay mode of the Higgs boson, one would think that it should be straightforward to measure it and hence determine the bottom quark Yukawa coupling $y_b$. However, reaching the current precision has proven to be a challenging endeavor.
The principal reason for this is that it is difficult to identify $b$ jets from Higgs boson decays over the much more numerous $b$ jets arising from quantum chromodynamics (QCD) backgrounds. It was pointed out in Ref.~\cite{Butterworth:2008iy} that requiring a boosted Higgs boson to be produced in association with either a weak vector boson or a jet could allow the $H \to b \bar{b}$ decay to be identified and studied using jet substructure techniques. Following this strategy, both the ATLAS and CMS experiments observed the $H \to b \bar{b}$ decay in associated $VH$ production~\cite{CMS:2017odg,CMS:2018nsn,ATLAS:2017cen,ATLAS:2018kot} and measured the bottom quark Yukawa coupling.

Another promising production mode which could lead to complementary studies of the $H \to b \bar{b}$ decay is weak boson fusion (WBF), $pp \to H(\to b \bar{b})jj$. In weak boson fusion, Higgs bosons are produced at central rapidities, together with two nearly back-to-back forward jets with little hadronic activity between them. This means that Higgs production in WBF followed by $H \to b \bar{b}$ decay can be distinguished from the QCD background by the presence of two central $b$ jets and two forward jets, typically originating from light quarks. Indeed, in the case of WBF with $H \to b \bar{b}$ decay, ATLAS and CMS report signal strengths consistent with the Standard Model and claim that observed rates are about $2.5$ standard deviations away from the background-only hypothesis~\cite{ATLAS:2020bhl,ATLAS:2020cvh,CMS:2023tfj,CMS:2023vid}. These measurements will be further improved during Run 3 and beyond.

In order to distinguish the signal process from the backgrounds and to extract interesting quantities from the signal measurement, it is imperative to have the best possible theoretical predictions for both. In this paper, we focus on the next-to-leading-order (NLO) and next-to-next-to-leading-order (NNLO) QCD corrections to the signal process $p p \to H(\to b \bar{b})jj$.
In the narrow width approximation, this process consists of on-shell Higgs boson production in weak boson fusion followed by the decay of the Higgs boson $H \to b \bar{b} + X$. Since the Higgs boson is a spin-zero particle, there are no spin correlations and the factorization between production and decay would be complete if not for selection criteria applied to QCD jets from both subprocesses.

Higgs boson production in WBF has been known through NNLO accuracy in QCD at a fully-differential level for about a decade~\cite{Cacciari:2015jma, Cruz-Martinez:2018rod}.\footnote{For earlier NLO QCD computations, see Refs.~\cite{Figy:2003nv,Berger:2004pca}.} Furthermore, the WBF cross sections that are inclusive with respect to QCD radiation have been computed with an astonishing N$^3$LO QCD accuracy~\cite{Dreyer:2016oyx}.
These theoretical results allow us to reduce the uncertainties in perturbative QCD predictions for fiducial cross sections of Higgs boson production in WBF to just a few percent.
Given this level of precision, one needs to critically examine other parametrically-suppressed contributions which were neglected earlier, but may become relevant at the achieved level of precision.
Such effects include electroweak corrections to Higgs production in WBF \cite{Ciccolini:2007jr,Ciccolini:2007ec}, interference effects in collisions of identical quarks \cite{Ciccolini:2007jr,Ciccolini:2007ec}, as well as nonfactorizable contributions that describe a cross talk between the two incoming quark legs and appear for the first time at NNLO in QCD~\cite{Liu:2019tuy,Dreyer:2020urf,Bronnum-Hansen:2023vzh,Long:2023mvc,Asteriadis:2023nyl}.
The results of these analyses lead to the conclusion that the effects mentioned above provide percent-level corrections to WBF fiducial cross sections, which should be included in the most up-to-date theoretical predictions for this quantity.

Interestingly, in spite of this enormous effort, the impact of Higgs boson decays on the magnitude of radiative corrections in weak boson fusion remained largely unexplored.
Understanding this is particularly important for the $H \to b \bar{b}$ decay channel because of the rich jet environment in this process and the selection criteria used to identify it.
While the WBF-production jets and the decay jets typically populate different regions of phase space, identifying $b$ jets requires them to pass a certain $p_\perp$-threshold. This selects events with particular values of the Higgs boson transverse momentum and impacts the magnitude of radiative corrections~\cite{Asteriadis:2021gpd}. Furthermore, the transverse momenta of identified $b$ jets are affected by the QCD radiation in the $H \to b \bar{b}$ decay, thus changing the fraction of decay products which pass kinematic cuts.

The first step towards a better understanding of the interplay between Higgs boson production and decay in WBF was made in Ref.~\cite{Asteriadis:2021gpd}, where both $H \to b \bar{b} $ and $H \to W^+W^- \to 4l$ decays were studied. It was found that imposing fiducial cuts on the Higgs decay products could change the NNLO QCD corrections by an order-one factor, making the impact of these cuts comparable to the size of the NNLO corrections themselves. In this reference, the WBF production process was considered at NNLO QCD, but the decay $H \to b \bar{b}$ was treated at LO only. In the present paper, we extend the study of Ref.~\cite{Asteriadis:2021gpd} and include QCD corrections to the Higgs decay as well, providing, for the first time, an NNLO-QCD-accurate description of the combined WBF process $pp \to H (\to b \bar{b}) jj$. As we will show in this paper, the magnitude of radiative corrections to $pp \to H (\to b \bar{b}) jj$ is significantly affected by the transverse momentum cut on $b$ jets. We find that for the values of this cut used in recent analyses by the ATLAS and CMS collaborations, the QCD corrections are significant, reducing the LO prediction by about $\qty{40}{\percent}$.

The remainder of the paper is organized as follows. In Section~\ref{sec:details} we summarize the salient technical details of our calculation. We explain how corrections to the $H \to b \bar{b}$ decay are included in the calculation and discuss the approximations we employ. We present phenomenological results in Section~\ref{sec:results} where we also investigate the origin of the fairly large QCD corrections that appear when Higgs decay to $b \bar{b}$ pairs is included in the calculation. We conclude in Section~\ref{sec:conclusions}.


\section{Technical Details}
\label{sec:details}

We consider Higgs boson production in weak boson fusion $pp \to Hjj$ followed by the decay $H \to b \bar{b}$. The combined process is described in the narrow width approximation, which implies that the production and decay processes factorize, the Higgs boson is always on the mass shell, and the QCD corrections to the production and decay can be considered separately.
We therefore write the differential cross section as
\begin{align}
    \textrm{d} \sigma =
    \textrm{Br}_{H \to b \bar{b}} \
    \textrm{d} \sigma_\textrm{WBF} \
    \textrm{d} \gamma_b \, ,
\label{eq1}
\end{align}
where $\textrm{Br}_{H \to b \bar{b}} $ is the $H \to b \bar b$ branching ratio, $\textrm{d} \sigma_\textrm{WBF}$ is the differential WBF production cross section, and ${\rm d}\gamma_b$ is the ratio of the differential and inclusive widths of the $H \to b \bar{b}$ decay
\begin{align}
{\rm d} \gamma_b = \frac{{\rm d} \Gamma_{b}}{\Gamma_b} \, .
\label{eq2}
\end{align}

In principle, all quantities shown on the right-hand side in Eq.~\eqref{eq1} receive QCD corrections and the perturbative expansion of ${\rm d} \sigma$ should involve all of them.
However, since our focus in this paper is on how QCD corrections to the decay can impact the fiducial cross section, we decide to set
${\rm Br}_{H \to b \bar{b}}$ to a fixed value,
${\rm Br}_{H \to b \bar{b}}=0.5824$
\cite{LHCHiggsCrossSectionWorkingGroup:2016ypw}.
At the same time, if a discussion of perturbative convergence of ${\rm d} \sigma $, including the perturbative expansion of ${\rm Br}_{H \to b \bar{b}}$, is desired, it can easily be constructed from the results that we present below.
The last reason for writing the differential cross section as in Eq.~\eqref{eq1} is that, for an observable that is inclusive in the decay products of the Higgs boson, the impact of the decay is captured by the branching ratio alone, because
\begin{align}
  \int {\rm d} \gamma_b = 1 \, .
\end{align}

We then proceed by separately expanding the quantities $\mathrm{d}\sigma_{\rm WBF}$, $\mathrm{d}\Gamma_b$, and $\Gamma_b$ in $\alpha_s$, following the prescription of Refs.~\cite{Ferrera:2013yga,Ferrera:2014lca,Campbell:2016jau,Ferrera:2017zex, Caola:2017xuq, Behring:2020uzq}. Thus, denoting the $i$-th term in the $\alpha_s$ expansion of a quantity $X$ as $X^{(i)}$, with the appropriate power of $\alpha_s$ included in $X^{(i)}$, we construct an $\text{N$^n$LO}$ approximation of $\rmd\sigma$
\begin{align}
	\mathrm{d}\sigma^{\text{N$^n$LO}}
	={\rm Br}_{H \to b \bar{b}}\;\sum\limits_{k=0}^n\;\mathrm{d}\sigma_{\rm WBF}^{(n-k)}\times\mathrm{d}\gamma^{\mathrm{N}^k\mathrm{LO}} \, ,
	\label{eq4}
\end{align}
where
\begin{align}
	\mathrm{d}\gamma^{\text{N$^n$LO}}
    =\frac{\sum_{k=0}^n\mathrm{d}\Gamma_b^{(k)}}{\Gamma_b^{\text{N$^n$LO}}}
	\label{eq5}
    \,, \quad
    \Gamma_b^{\text{N$^n$LO}} = \sum_{k=0}^n \Gamma_b^{(k)}
    \,.
\end{align}
We note that the quantities $\mathrm{d}\gamma^\text{N$^n$LO}$, integrated over fiducial phase space, describe a mismatch between corrections to inclusive and fiducial decay widths and, therefore, are very important for the discussion of the impact of QCD corrections to $H \to b \bar{b}$ decay on Higgs production in WBF in the presence of kinematic cuts on final-state particles.

Furthermore, although it would be legitimate to expand the inverse width in the definition of $\mathrm{d}\gamma^\text{N$^n$LO}$ in a series in $\alpha_s$, we decided not to do so because the NLO corrections to $\Gamma_{b}$ are fairly large. Clearly, our results for $\mathrm{d}\sigma^{\text{N$^n$LO}}$ are accurate up to and including $\mathcal{O}(\alpha_s^n)$.
We will further discuss this and other alternatives for providing an NNLO prediction for the fiducial cross section in Section~\ref{sec:results}.

The calculation of $\mathrm{d}\sigma^\text{N$^n$LO}$, $n = 0,1,2$, does not require any new theoretical computation since fully-differential descriptions of both ${\rm d} \sigma_{\rm WBF}$ and ${\rm d} \gamma$ through NNLO in perturbative QCD are available~\cite{Asteriadis:2021gpd,Behring:2019oci}. The latter computation was performed in the Higgs boson rest frame but, since the Higgs is a spin-zero particle, this is sufficient.
We obtain the Higgs boson momentum from production kinematics, let it decay in its rest frame and then boost the momenta of all particles that originate from Higgs decay to the laboratory frame where the WBF production process occurs.

We emphasize that, although the step of combining theoretical predictions for Higgs production and decay in WBF is simple conceptually, doing so in practice requires care. Indeed, we need to interface the code for computing NNLO QCD corrections to the production described in Ref.~\cite{Asteriadis:2021gpd} and the code for computing NNLO QCD corrections to $H \to b \bar{b}$ decay presented in Ref.~\cite{Behring:2019oci}.
Both codes produce fully-differential results for the respective quantities through NNLO in perturbative QCD using local subtraction schemes. As a result, each subprocess comprises terms with varying multiplicities, weights and four-momenta. Hence, we need to ensure that these quantities are properly combined such that selection criteria can be applied to the complete set of final-state particles in an infrared-safe way.

Another important aspect that needs to be addressed is the efficiency of the interfaced code. We have pointed out in Ref.~\cite{Asteriadis:2021gpd} that the complexity of the NNLO QCD computation in the WBF process makes it difficult to consider complex decay products of the Higgs boson because the already-large dimensionality of the phase space becomes even larger. Since for processes of the DIS-type---to which Higgs production in WBF belongs---the adaptation of the integration grids in higher-order computations may be a challenge, any increase in the dimensionality of the phase space may have a detrimental effect on the efficiency of the computation.
In the current case, we have solved these problems by preparing dedicated integration grids for different parts of the calculation, using additional parameters to change the magnitude of certain subtraction terms, working with MPI protocols to parallelize the computation, and simply using very large samples of points for Monte-Carlo integration. Nevertheless, even with all these improvements, computational resources of roughly $\num{5e5}$ CPU hours were required to produce the numerical results that we present in the next section.

Finally, since Higgs decays to $b$ quarks in the WBF process are identified by the presence of two $b$ jets in the final state, and since the identification of the flavor of jets in perturbative computations is subtle \cite{Banfi:2006hf}, it is important to note that the calculation of Ref.~\cite{Behring:2019oci} is performed for \emph{massive} $b$ quarks.
This has important consequences because it allows us to directly employ the standard anti-$k_\perp$ algorithm~\cite{Cacciari:2008gp,Cacciari:2011ma} to identify $b$ jets in an infrared-safe manner.\footnote{As an alternative, we could perform the calculation using massless $b$ quarks in the decay and either the flavor-$k_\perp$ algorithm~\cite{Banfi:2006hf} or one of the new-generation flavor-safe variants of the anti-$k_\perp$ algorithm~\cite{Caletti:2022hnc, Czakon:2022wam,Gauld:2022lem,Caola:2023wpj}.}
However, we note that Higgs boson production in WBF is still computed with \emph{massless} $b$ quarks. We ensure infrared and collinear safety of our calculation by only considering $b$ quarks from the decay subprocess in the flavor tagging, effectively treating $b$ quarks originating from the production as flavorless. This makes our computation incomplete, and in order to complete it, one should also use massive quarks in the production process. However, as we argue below, at the end of Section~\ref{sec:results}, the omitted effects are minor compared to the fiducial volume effects introduced by the inclusion of $H \to b \bar{b}$ decays in the theoretical predictions.


\section{Results}
\label{sec:results}

For numerical computations, we employ the standard set-up that has been used to provide theoretical descriptions of weak boson fusion in the past \cite{Asteriadis:2021gpd,Cacciari:2015jma,Cruz-Martinez:2018rod}. We consider proton-proton collisions at \qty{13}{\TeV} center-of-mass energy.
We set the mass of the Higgs boson to
$m_H = \qty{125}{\GeV}$,
the mass of the $W$ boson to $m_W=\qty{80.398}{\GeV}$ and the mass of the $Z$ boson to $m_Z = \qty{91.1876}{\GeV}$.
As is customary, we include widths of electroweak gauge bosons in their propagators and choose $\Gamma_W = \qty{2.1054}{\GeV}$ and $\Gamma_Z = \qty{2.4952}{\GeV}$ for the numerical values, respectively.
Weak couplings are derived from the Fermi constant $G_F = \qty{1.16639e-5}{\GeV^{-2}}$ and the CKM matrix is set to the identity matrix.\footnote{This does not affect the predictions because of CKM-unitarity and the fact that we treat all the light jets in the production on equal footing.}

The $b$ quark Yukawa coupling $\overline{y}_b$ is computed in the $\overline{\text{MS}}$ scheme \cite{Braaten:1980yq} using the $b$-quark mass $\overline{m}_b(\mu=m_H)=\qty{2.81}{\GeV}$ and the relation $\overline{y}_b = \overline{m}_b (2 \sqrt{2} G_F)^{1/2}$. We use the package \textsc{RunDec}~\cite{Chetyrkin:2000yt,Schmidt:2012az,Herren:2017osy} to calculate the running at two-loop order using five active quark flavors. For the $b$-quark mass that appears in the matrix elements and in the phase-space of $H \to b \bar{b}$ decay, we use the pole mass with a numerical value $m_b^\text{pole}=\qty{4.78}{\GeV}$. Furthermore, at NNLO, there is a contribution to $H \to b \bar{b}$ decays which is mediated by a top-quark loop \cite{Larin:1995sq,Chetyrkin:1995pd,Bernreuther:2018ynm,Primo:2018zby,Behring:2019oci}. To compute this contribution we use $m_t^\text{pole}=\qty{173.2}{\GeV}$; the top quark Yukawa coupling is derived from this value as well, via $y_t = m_t^\text{pole} (2 \sqrt{2} G_F)^{1/2}$.

With this setup, the $H \to b\bar{b}$ decay width for $\mu=m_H$ at different orders is
\begin{align}
\begin{aligned}
    \Gamma_b^{\rm LO} &=\qty{1.926}{\MeV} \, , \\
    \Gamma_b^{\rm NLO} &=\qty{2.327}{\MeV} \, , \\
    \Gamma_b^{\rm NNLO} &=\qty{2.432}{\MeV} \, .
    \label{decay_width}
\end{aligned}
\end{align}
The Monte Carlo integration uncertainties in the decay rate $\Gamma_b$ are negligible in comparison to the Monte Carlo uncertainty in the fiducial cross-section $\sigma$.
Although these decay widths were computed using the same code as was used in Ref.~\cite{Behring:2019oci}, they are slightly different from the results presented there due to somewhat different choices of input parameters, most notably the top quark Yukawa coupling.

We employ \texttt{NNPDF31\_nnlo\_as\_0118} parton distribution functions~\cite{NNPDF:2017mvq} and $\alpha_s(m_Z) = 0.118$ for all calculations reported below. The evolution of both parton distribution functions
and the strong coupling constant is obtained directly from LHAPDF~\cite{Buckley:2014ana}.
We use equal renormalization and factorization scales to describe Higgs production in WBF.
As a central scale, we choose~\cite{Cacciari:2015jma}
\begin{align}
  \mu_0^2 = \frac{m_H}{2} \sqrt{\frac{m_H^2}{4} + p_{\perp,H}^2}
  \, ,
\end{align}
where $p_{\perp,H}$ is the \emph{true} Higgs boson transverse momentum.
To study the dependence of theoretical predictions on the renormalization and factorization scales, we compute them for $\mu_R = \mu_F = 2 \mu_0$ and $\mu_R = \mu_F = \mu_0/2$.
We choose $\mu = m_H$ as our central scale for the Higgs boson decay subprocess. We will not investigate the impact of the scale variation in the decay systematically, but we will make some comments about how such a choice impacts theoretical predictions.

Final-state jets, including $b$ jets, are defined using the standard anti-$k_\perp$ algorithm~\cite{Cacciari:2008gp,Cacciari:2011ma} with a jet radius parameter of $R = 0.4$.
We select events that have at least two $b$ jets in the final state, where, following the ATLAS analysis in Ref.~\cite{ATLAS:2018kot}, each of the $b$ jets is required to have a transverse momentum $p_{\perp,b} > \qty{65}{\GeV}$ and a rapidity $|y_b| < 2.5$.
We remind the reader that in our calculation $b$ jets come \emph{exclusively} from Higgs decays.
Most of our events have, therefore, at most two $b$ jets, although signatures with four $b$ jets may arise due to the NNLO corrections to the decay.
In this case, we consider the two $b$ jets with an invariant mass that is closest to the mass of the Higgs boson as a candidate pair to reconstruct the Higgs boson momentum.
If more than one $b$ and/or $\bar b$ quark are clustered into a jet, we still call it a $b$ jet and include it in the $b$-jet counting.
It is clear that there is a certain ambiguity in how such multiple $b$ tags should be handled, but the whole effect is small and does not change the overall pattern of radiative corrections that we describe below.

Once two $b$ jets that reconstruct the Higgs boson are identified, we apply standard WBF cuts to the remaining jets, which we will refer to as ``light jets''.\footnote{Note that the light jets may arise from radiative corrections to either the production or decay process.}
Specifically, we require that a WBF event contains at least two light jets with transverse momenta $p_{\perp,j} > \qty{25}{\GeV}$ and rapidities $|y_j| < 4.5$.
The two leading-$p_\perp$ light jets must have well-separated rapidities, $|y_{j_1} - y_{j_2}| > 4.5$, and their invariant mass should be at least \qty{600}{\GeV}.
In addition, the two leading light jets must be in opposite hemispheres in the laboratory frame; this is enforced by requiring that the product of their rapidities in the laboratory frame is negative, $y_{j_1} y_{j_2} < 0$.

We are now in the position to present the results of the calculation. We start by computing fiducial cross sections at LO, NLO and NNLO in the perturbative QCD expansion.
We find\footnote{The small difference in the LO cross section compared to the result reported in Ref.~\cite{Asteriadis:2021gpd} is due to the inclusion of the nonzero $b$-quark mass in $H \to b\bar{b}$ decays in this computation.}
\begin{align}
\begin{split}
    \sigma^{\rm   LO} = \qty[parse-numbers=false]{75.6_{+6.5}^{-5.6}}{\femto\barn} \, , \\
    \sigma^{\rm  NLO} = \qty[parse-numbers=false]{52.4_{-2.6}^{+1.5}}{\femto\barn} \, , \\
    \sigma^{\rm NNLO} = \qty[parse-numbers=false]{44.6_{-0.6}^{+0.9}}{\femto\barn} \, ,
\label{eq7}
\end{split}
\end{align}
where the superscripts and subscripts indicate values with $\mu_R = \mu_F = 2\mu_0$ and $\mu_R = \mu_F = \mu_0/2$, respectively.
Monte Carlo integration uncertainties are about a few permille and thus much smaller than the residual scale uncertainties; consequently, we do not show them.
The results in Eq.~\eqref{eq7} show that, compared to the LO cross section, the NLO cross section is reduced by about $\qty{30}{\percent}$ and the NNLO cross section by an additional $\qty{10}{\percent}$. These significant corrections should be contrasted with the scale uncertainties, which are about $\qty{9}{\percent}$, $\qty{5}{\percent}$, and $\qty{2}{\percent}$ on the LO, NLO and NNLO predictions, respectively.
Hence, in this case, the strategy of estimating uncertainties of theoretical predictions by studying their sensitivity to renormalization and factorization scales does not capture the contributions from missing higher orders.

The large negative QCD corrections seen in Eq.~\eqref{eq7} arise because corrections to the decay in the fiducial region are different from those to the inclusive width and, while they are moderate in both cases, they conspire to reduce the fiducial cross section.
We will now illustrate this statement by considering the different contributions that lead to the results shown in Eq.~\eqref{eq7}.

\begin{figure*}[t]
    \centering
    \includegraphics[height=280pt]{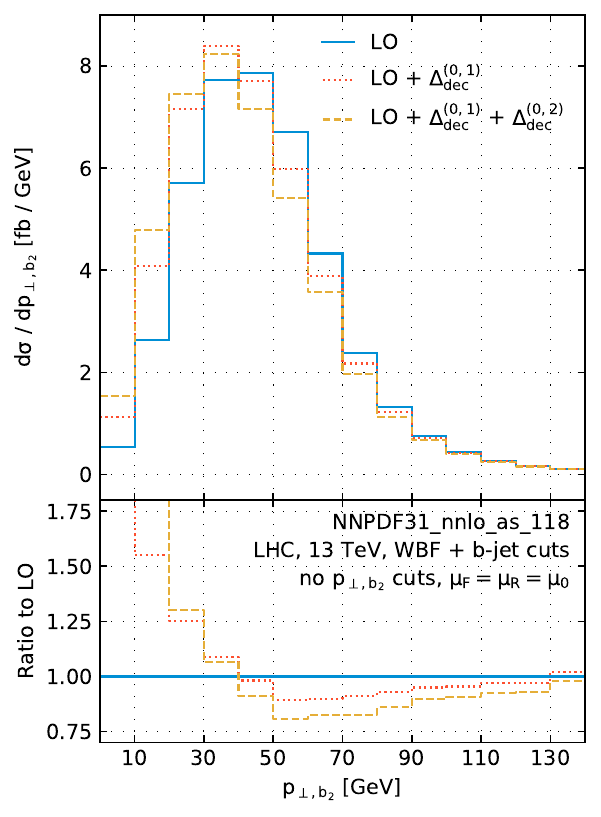}
    \hspace{40pt}
    \includegraphics[height=280pt]{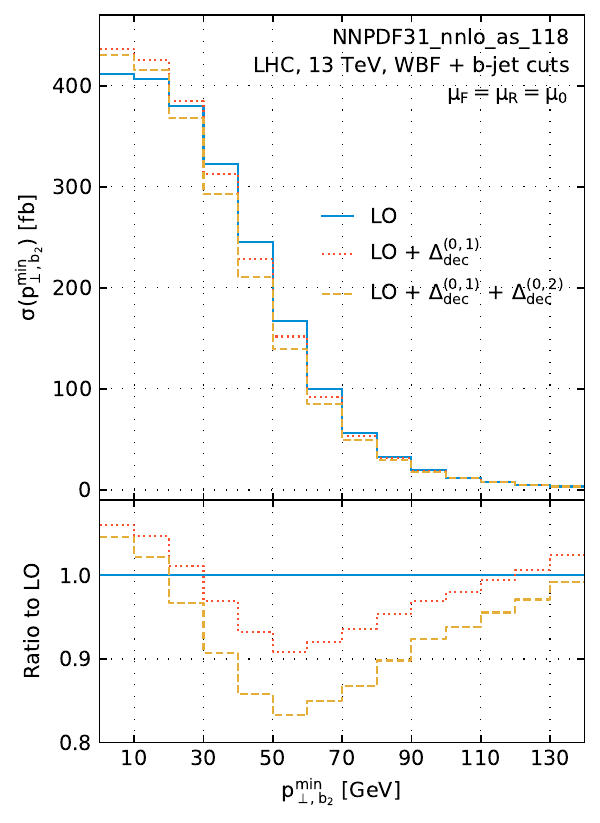}
    \hspace{10pt}
    \caption{The impact of QCD corrections to the $H\to b \bar{b}$ decay on the transverse momentum distribution of the subleading $b$ jet (left) and on the fiducial cross section defined in Eq.~\eqref{cumulative-xs} (right). In each plot, the upper pane shows results at LO, NLO and NNLO in the decay process while the Higgs production process is kept at LO QCD. The lower panes show the ratio with respect to the LO results. See text for details.}
    \label{fig:ptb2_nocut}
\end{figure*}

It follows from Eqs.~\eqref{eq4}~and~\eqref{eq5} that higher-order QCD corrections to the differential WBF cross section with $H \to b \bar{b}$ decay
originate from the interplay of three different contributions: the radiative corrections to the WBF production process ${\rm d} \sigma_{\rm WBF}$, the radiative corrections to the differential decay process ${\rm d} \Gamma_b$, and the expansion of the total $H \to b \bar{b}$ width $\Gamma_b$ in $\alpha_s$. We will refer to these corrections $\Delprod$, $\Deldec$ and $\Delexp$, respectively.
Beyond LO we write the cross section Eq.~\eqref{eq4} as
\begin{align}
{\rm d} \sigma^\text{N$^n$LO}
= {\rm d} \sigma^\mathrm{LO}
+ \sum_{k = 1}^n {\rm d} \Delta^{\mathrm{N}^k\mathrm{LO}} \, .
\end{align}
We write
\begin{align}
\begin{split}
{\rm d} \Delta^{\rm NLO} ={}&
\mathrm{d}\Delprod^{(1,0)}
+\mathrm{d}\Deldec^{(0,1)}
+ \mathrm{d}\Delexp^{(0,1)}
\,, \\
\rmd\Delta^{\rm NNLO} ={}&
\rmd\Delprod^{(2,0)}
+\rmd\Deldec^{(1,1)}
+\rmd\Deldec^{(0,2)}
 \\ &
+\rmd\Delexp^{(1,1)}
+\rmd\Delexp^{(0,2)}
\,.
\end{split}
\end{align}
The individual contributions are defined as
\begin{align}
	\begin{split}
\mathrm{d}\Delprod^{(1,0)}&=\hphantom{-}\textrm{Br}_{H \to b\bar{b}} \; \bigg(\mathrm{d}\sigma_{\rm WBF}^{(1)} \times \frac{\mathrm{d}\Gamma_b^{(0)}}{\Gamma_b^\mathrm{(0)}} \bigg) \,,\\
	\mathrm{d}\Deldec^{(0,1)}&=\hphantom{-}\textrm{Br}_{H\to b\bar{b}}\;\frac{\Gamma_b^{(1)}}{\Gamma_b^\mathrm{NLO}} \bigg( \mathrm{d}\sigma_{\rm WBF}^{(0)} \times \frac{\mathrm{d}\Gamma_b^{(1)} }{\Gamma_b^{(1)}} \bigg) \,,\\
	\mathrm{d}\Delexp^{(0,1)}&=-\textrm{Br}_{H\to b\bar{b}}\;\frac{\Gamma_b^{(1)}}{\Gamma_b^\mathrm{NLO}}\bigg(\mathrm{d}\sigma_{\rm WBF}^{(0)}\times \mathrm{d}\gamma^\mathrm{LO}\bigg)\,,
	\end{split}
\end{align}
and
\begin{align}
	\mathrm{d}\Delprod^{(2,0)}&=\hphantom{-}\textrm{Br}_{H\to b\bar{b}} \; \bigg(\mathrm{d}\sigma_{\rm WBF}^{(2)} \times \frac{\mathrm{d}\Gamma_b^{(0)}}{\Gamma_b^{(0)}} \bigg) \,, \nonumber \\
	\mathrm{d}\Deldec^{(1,1)}&=\hphantom{-}\textrm{Br}_{H\to b\bar{b}}\; \frac{\Gamma_b^{(1)}}{\Gamma_b^\mathrm{NLO}} \bigg( \mathrm{d}\sigma_{\rm WBF}^{(1)} \times \frac{\mathrm{d}\Gamma_b^{(1)}}{\Gamma_b^{(1)}} \bigg)\,, \nonumber \\
	\mathrm{d}\Deldec^{(0,2)}&=\hphantom{-}\textrm{Br}_{H\to b\bar{b}}\; \frac{\Gamma_b^{(2)}}{\Gamma_b^\mathrm{NNLO}} \bigg(\mathrm{d}\sigma_{\rm WBF}^{(0)} \times \frac{\mathrm{d}\Gamma_b^{(2)}}{\Gamma_b^{(2)}} \bigg) \,,\\
	\mathrm{d}\Delexp^{(1,1)}&=-\textrm{Br}_{H\to b\bar{b}}\;\frac{\Gamma_b^{(1)}}{\Gamma_b^\mathrm{NLO}} \bigg(\mathrm{d}\sigma_{\rm WBF}^{(1)} \times \mathrm{d}\gamma^\mathrm{LO} \bigg) \,, \nonumber \\
	\mathrm{d}\Delexp^{(0,2)}&=-\textrm{Br}_{H\to b\bar{b}}\;\frac{\Gamma_b^{(2)}}{\Gamma_b^\mathrm{NNLO}} \bigg( \mathrm{d}\sigma_{\rm WBF}^{(0)} \times \mathrm{d}\gamma^\mathrm{NLO} \bigg) \,. \nonumber
\end{align}
We choose $\mu_{\rm R} = \mu_{\rm F} = \mu_0$, integrate over the fiducial phase space, and find
\begin{align}
\sisetup{retain-explicit-plus=true}
\begin{split}
    \Delprod^{(1,0)} &= \qty{-4.9}{\femto\barn}\, , \quad
    \Deldec^{(0,1)} = \qty{-5.3}{\femto\barn}\, , \\
    \Delexp^{(0,1)} &= \qty{-13.0}{\femto\barn} \, , \\
    \Delprod^{(2,0)} &= \qty{-1.5}{\femto\barn} \, , \quad
    \Deldec^{(1,1)} = \qty{+0.4}{\femto\barn} \, , \\
    \Deldec^{(0,2)} &= \qty{-5.0}{\femto\barn}\, , \quad
    \Delexp^{(1,1)} = \qty{+0.8}{\femto\barn} \, , \\
    \Delexp^{(0,2)} &= \qty{-2.5}{\femto\barn} \, .
\end{split}
\label{eq9}
\end{align}
It follows from Eqs.~\eqref{eq7}~and~\eqref{eq9} that the corrections arising from the production process ($\Delta_{\rm prod}^{(1,0)}$ and $\Delta_{\rm prod}^{(2,0)}$) are small and negative;
they amount to approximately $\qty{-6}{\percent}$ at NLO and $\qty{-2}{\percent}$ at NNLO, and hence are covered by scale uncertainty estimates.
These corrections were extensively discussed in Ref.~\cite{Asteriadis:2021gpd}.
The corrections associated with the differential decay process ($\Deldec^{(0,1)}$ and $\Deldec^{(0,2)}$)
are both about $\qty{-7}{\percent}$ of the LO cross section, indicating that the perturbative expansion does not seem to converge.
Moreover, since they are also negative, they amplify corrections to the production.
We study these rather large corrections in more detail below.

Before doing so, we note that the corrections arising from the expansion of the width $\Gamma_b$,
which appears in the definition of the cross section through Eq.~\eqref{eq5},
are significant and \emph{also} negative.
At NLO they lead to a large decrease in the fiducial cross section shown in Eq.~\eqref{eq7}.
At NNLO, however, the corrections from expanding $\Gamma_b$ are negative but moderate, giving an overall decrease of the cross section by about $\qty{2}{\percent}$ with respect to the LO value.
Thus, we conclude that the reason for large
QCD corrections to the WBF process with Higgs decay to
a $b \bar{b}$ pair is the fact that a number of moderately-large
negative corrections amplify each other and lead to a significant decrease in the fiducial cross section when taken together.

The feature of NNLO QCD corrections that clearly stands out is the lack of apparent convergence of the perturbative expansion of the decay width $H \to b \bar{b}$ in the fiducial volume.
To understand this feature better, we note that the cut on the transverse momenta of the two $b$ jets, $p_{\perp,b} \ge \qty{65}{\GeV}$, is rather high as it exceeds $m_H/2$. Hence, if the Higgs bosons were produced with zero transverse momentum, none of the $b$ jets from Higgs boson decays would pass this cut. Of course, this does not happen in WBF since the typical transverse momentum of Higgs bosons in this process is $p_{\perp,H} \approx \qty{120}{\GeV}$. Because of this, it may appear that the cut on the $b$-jet transverse momentum is sufficiently low to leave a fiducial phase space that is large enough for perturbative stability. However, as we will show below, this is not the case and this cut appears to be too restrictive from the point of view of perturbative convergence.

To illustrate this point, we recompute the decay corrections $\Deldec^{(0,i)}$, $i=1,2$, imposing all fiducial cuts \emph{except} the transverse momentum cut on the subleading $b$ jet, $p_{\perp,b_2}^\text{min}$, and study the transverse momentum distribution of this jet up to NNLO.
The results are shown in Fig.~\ref{fig:ptb2_nocut}.
The distribution peaks around $p_{\perp,b_2} \approx \qty{30}{\GeV}$ and then decreases quite rapidly between $40$ and $\qty{80}{\GeV}$.
The additional real emissions in the process $H \to b \bar{b}$ tend to reduce the transverse momentum of $b$ jets from Higgs decay, making their transverse momentum distribution softer. QCD corrections to the \emph{right} of the peak converge poorly,
with NLO and NNLO contributions being moderate but quite comparable without clear hierarchy. Only to the left of the peak at $p_{\perp,b_2} \approx \qty{30}{\GeV}$ do the corrections start exhibiting signs of perturbative convergence.

\begin{figure*}[t]
    \centering
    \includegraphics[height=270pt]{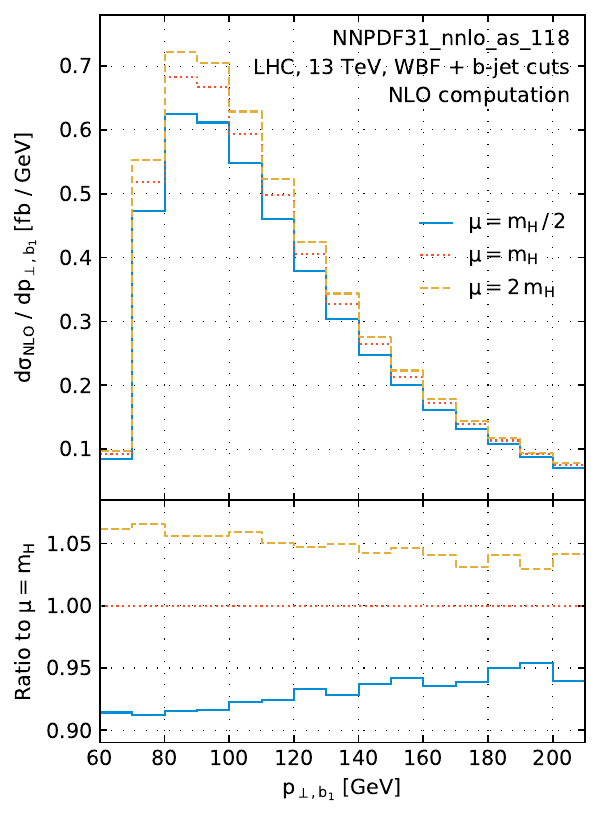}
    \hspace{55pt}
    \includegraphics[height=270pt]{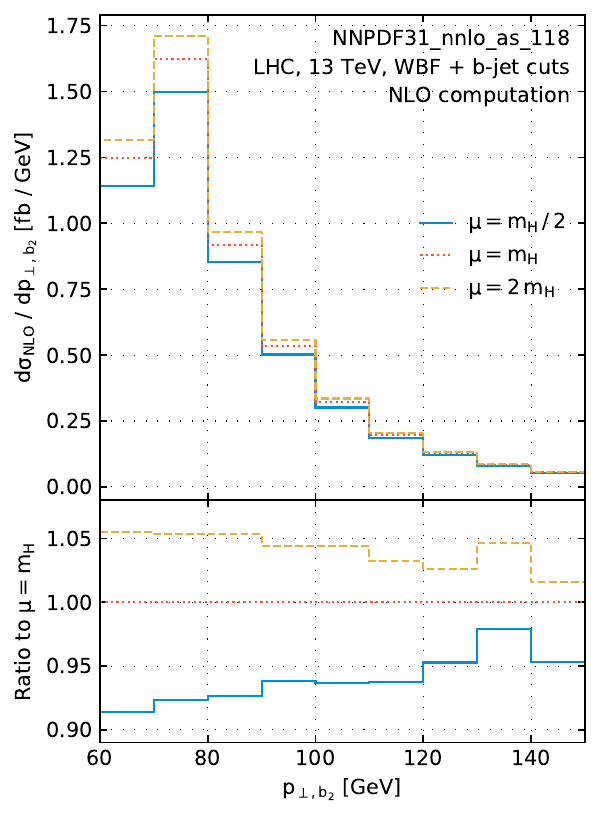}
    \hspace{10pt}
    \caption{Distributions of transverse momenta of the leading (left) and subleading (right) $b$ jets. In each plot, the upper pane shows the NLO computation for different renormalization scales used in the computation of the $H\to b\bar{b}$ decay width. In all these results the central scale choice is used for the renormalization and factorization scale in the Higgs production process. The lower pane shows the ratio to the central scale choice $\mu = m_H$. See text for details.}
    \label{fig:ptb_decay_scale_variations}
\end{figure*}

To emphasize this point, we show the cumulative distribution, i.e.
\begin{align}
\sigma(p_{\perp,{b_2}}^\textrm{min}) = \int_{p_{\perp,{b_2}}^\textrm{min}}^\infty \textrm{d}p_{\perp,{b_2}} \ \frac{\textrm{d}\sigma}{\textrm{d}p_{\perp,{b_2}}} \, ,
  \label{cumulative-xs}
\end{align}
in the right pane in Fig.~\ref{fig:ptb2_nocut}, from which one can infer the dependence of the fiducial cross section on the lower cut on $p_{\perp,{b_2}}$.
We emphasize that only corrections to the decay are included in the cross sections shown in that figure.
We observe that for $p^\textrm{min}_{\perp,b_2} \gtrsim \qty{30}{\GeV}$ the NLO and NNLO corrections are comparable in size and the perturbative convergence is quite poor; QCD corrections to the fiducial decay width start converging only for smaller values of this cut.

Since it is clear that the cut on $p_{\perp, b}$ plays an important role in separating the $H \to b \bar b$ signal for the background, relaxing it dramatically is not an option. On the other hand, we observe that the cut on the subleading $b$ jet is primarily responsible for the poor perturbative convergence. This suggests that it might be possible to keep a stringent cut on $p_{\perp, b_1}$ while relaxing that on $p_{\perp, b_2}$, allowing the discrimination of signal and background without ruining perturbative convergence. Of course, this would require detailed studies, which are beyond the scope of this paper.

As we already mentioned, the second source of large corrections comes from the expansion of the inclusive partial decay width $\Gamma_b$ in the denominator in Eq.~\eqref{eq2};
the NLO correction from this expansion changes the fiducial cross section by $\qty{-17}{\percent}$ in comparison to LO.
This is a consequence of the relatively large corrections to the partial width, cf. Eq.~\eqref{decay_width}.

\begin{figure*}[t]
    \centering
    \includegraphics[height=270pt]{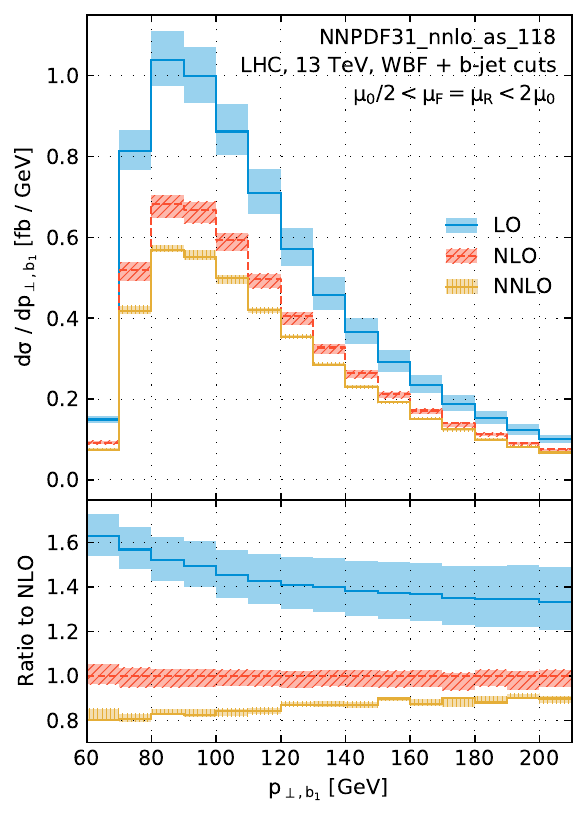}
    \hspace{55pt}
    \includegraphics[height=270pt]{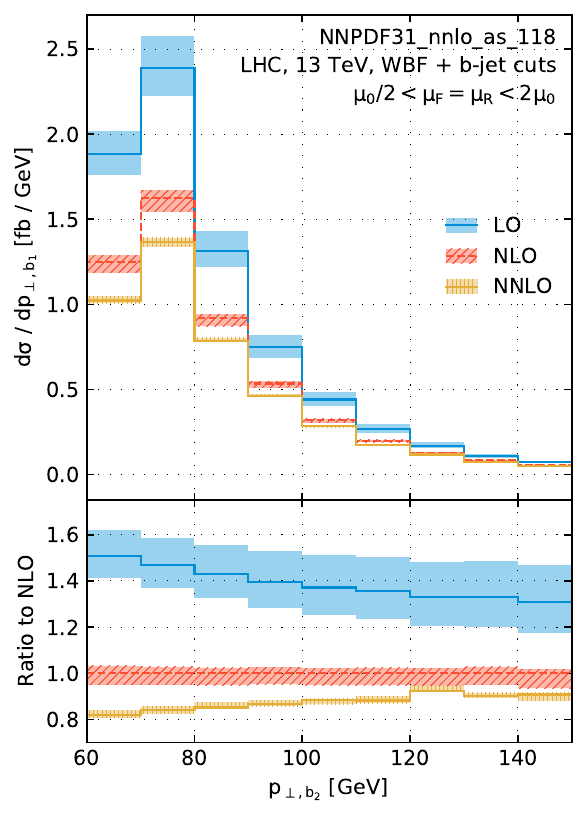}
    \hspace{10pt}
    \caption{Transverse momentum distributions of the leading (left) and subleading (right) $b$ jets. In each plot, the upper pane shows the LO, NLO, and NNLO QCD computations. The lines indicate the central renormalization and factorization scale choice, while the envelopes around them indicate the range of results for different scales in the production process. The lower panes show the ratios with respect to the NLO results at central scale. See text for details.}
    \label{figptb}
\end{figure*}

It was observed in Ref.~\cite{Behring:2019oci} that corrections to the width converge somewhat faster if the renormalization scale $\mu = m_H/2$ for the decay is chosen.
However, we do not see a clear benefit for using this scale in the WBF computation.
Indeed, for $\mu = m_H/2$, we find $\Delexp^{(0,1)}=\qty{-8.3}{\femto\barn}$ and $\Deldec^{(0,1)}=\qty{-13.8}{\femto\barn}$.
The two corrections add up to $\qty{-22.1}{\femto\barn}$ which should be compared to a similar result of $\qty{-18.3}{\femto\barn}$ for $\mu = m_H$.\footnote{We note that $\rmd\gamma^\text{LO}$, $\rmd\sigma^\text{LO}$, and $\rmd\Delta_{\rm prod}^{(1,0)}$ are independent of the renormalization scale in the decay.}
Hence, for $\mu = m_H/2$, corrections to ${\rm d} \sigma$ are somewhat larger than for $\mu = m_H$ and, more generally, the impact of the scale variation in the decay is comparable to that in the production.
The same conclusions can be drawn from the impact of decay scale variations on differential distributions. This statement is illustrated in Fig.~\ref{fig:ptb_decay_scale_variations}, where only NLO results are shown. We do this to verify that decay scale variations cannot capture the observed large shifts due to higher-order QCD corrections; because of that, we see no point in further pursuing detailed studies of the scale variation in the decay.

As noted earlier, there are different ways of combining the branching ratio ${\rm Br}_{H \to b \bar{b}}$ with the quantity ${\rm d} \sigma_{\rm WBF} \times {\rm d} \gamma$. In doing so, we need to remember that the formula in Eq.~(\ref{eq1}) is obtained by rewriting ${\rm d} \Gamma_b/\Gamma_H$ as the product of ${\rm Br}_{H \to b \bar{b}}$ and ${\rm d} \gamma = {\rm d} \Gamma_b/\Gamma_b$, and that the total width $\Gamma_b$ also appears in the numerator of ${\rm Br}_{H \to b \bar{b}}$. Hence, if we chose to expand ${\rm Br}_{H \to b \bar{b}}$ in $\alpha_s$, we must ensure that the total width $\Gamma_b$ is treated in the same way in both ${\rm Br}_{H \to b \bar{b}}$ and ${\rm d} \gamma$. For the prescription that we employ in this paper, this treatment works such that the quantity $\rmd\sigma^\mathrm{NNLO}$ is accurate up to and including $\mathcal{O}(\alpha_s^2)$. We note that other ways to construct the NNLO value for ${\rm d} \sigma$, for example expanding the inverse width $1/\Gamma_b$ in $\alpha_s$ or the prescription discussed in Ref.~\cite{Gauld:2019yng}, result in predictions for cross sections that differ from the NNLO result shown in Eq.~(\ref{eq7}) by about $\qty{2}{\femto\barn}$.

We continue with the discussion of QCD corrections to kinematic distributions in the WBF process for fiducial cuts (including $p_{\perp,b_2} > \qty{65}{\GeV}$).
We begin with displaying the transverse momentum of the leading and subleading $b$ jets in Fig.~\ref{figptb}.
The main effect of the higher-order QCD corrections on these distributions is a nearly uniform reduction of the LO results by an amount that is comparable to that of the fiducial cross section. This can be clearly seen from the ratio to NLO shown in the lower panes.
The NLO K-factors increase slightly from around $1/1.6\approx0.6$ for low values of $p_{\perp}$ to around $1/1.3\approx0.8$ in the tails of the distributions. No significant modifications of differential $K$-factors at NNLO are observed, so that the shape changes of these distributions are well-captured by NLO predictions.\footnote{We note that similar to the fiducial cross sections discussed in Eq.~(\ref{eq7}), the scale-variation bands for transverse momentum distributions in consecutive perturbative orders do not overlap either.}

\begin{figure*}[t]
    \centering
    \includegraphics[height=270pt]{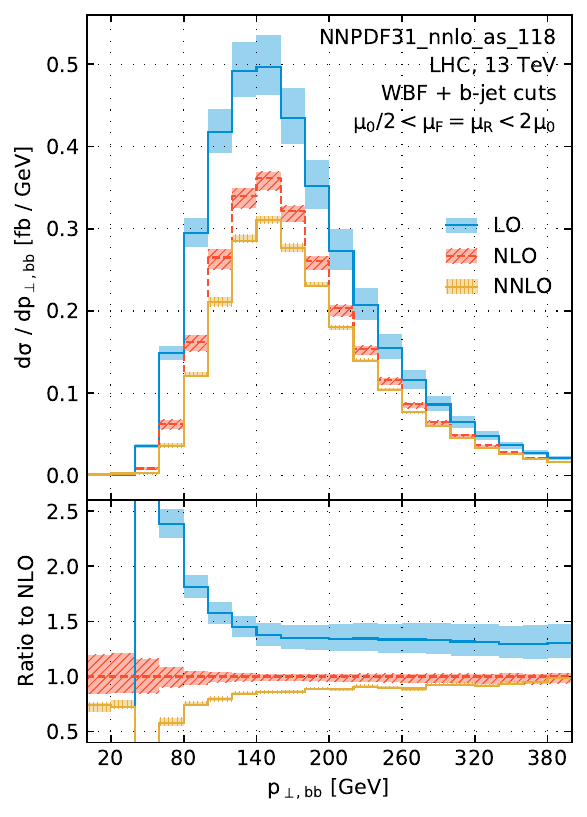}
    \hspace{55pt}
    \includegraphics[height=270pt]{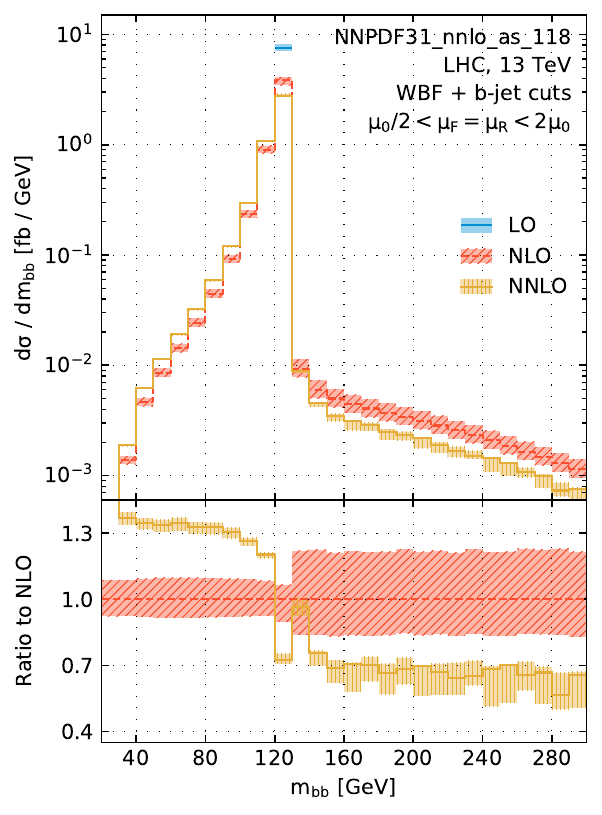}
    \hspace{10pt}
    \caption{ Distributions of the transverse momentum (left) and the invariant mass of the reconstructed Higgs boson (right). See caption of Fig.~\ref{figptb} and text for details.}
    \label{fig:reconH}
\end{figure*}

The situation changes, however, if we consider the transverse momentum distributions of the two $b$ jets whose invariant mass is closest to the Higgs boson mass, which can be interpreted as the transverse momentum of the ``reconstructed'' Higgs boson.
This distribution is shown in the left pane in Fig.~\ref{fig:reconH}.
We observe that in this case both the NLO and NNLO $K$-factors are rapidly-changing functions of the transverse momentum, with the NLO K-factor taking a value of around $1/2.4\approx0.4$ at $p_{\perp,bb} = \qty{70}{\GeV}$, rising rapidly to about $1/1.4\approx0.7$ at $p_{\perp,bb} = \qty{150}{\GeV}$ and leveling off after that.
We also observe a more significant difference between NNLO and NLO distributions as compared to the $p_\perp$-distributions of individual jets. This behavior is caused by a strong correlation between the value of the (true) Higgs boson transverse momentum and the ability of the produced $b$ jets to pass the cuts, especially when additional QCD radiation is considered.

Another distribution that is interesting to discuss is the invariant mass distribution of the two reconstructed $b$ jets, $m_{bb}$; it is shown in the right pane of Fig.~\ref{fig:reconH}.
In the absence of QCD radiation in the decay and when no interplay between production and decay is considered, this distribution populates a single bin because the momenta of two $b$ jets always reconstruct the mass of the Higgs boson. However, when radiative corrections are considered and final-state partons from the production process and Higgs decay are treated on an equal footing, the invariant mass distribution of the two $b$ jets is altered significantly.

As follows from the right pane in Fig.~\ref{fig:reconH}, there is a tail to the left of $m_{bb} = m_H$, caused by the QCD radiation in the decay, and a tail to the right, caused by the fact that occasionally the QCD radiation in production and one of the $b$ quarks from $H \to b\bar{b}$ decay are combined. Note that the ``left'' tail exhibits large positive NNLO contributions as more and more events populate this region.

This plot allows us to indirectly address the issue of whether it is justifiable to disregard $b$ jets from the production.
We can estimate the significance of clustering partons from production and partons from decays into the $b$ jets by counting the number of events with an invariant mass $m_{bb}$ above $m_H$ and comparing it to the total number of Higgs bosons produced.
We find that the number of such events is about $\qty{1}{\percent}$ of the total and that it is not drastically affected by the NNLO QCD corrections.
We may interpret this result as a measure of how often one of the $b$ quarks from Higgs decay and \emph{any} of the partons from the production find themselves sufficiently close to each other to be clustered into a jet. Since in this article we did not consider production of $b$ jets in the weak boson fusion process proper, this result gives us an estimate of how often partons from the production process end up in the ``decay'' fiducial region.

Furthermore, by considering simple LO processes, we have estimated that additional $b$ jets are present in roughly $\qty{1}{\percent}$ of events that contribute to the fiducial WBF cross sections in Eq.~\eqref{eq7}. This is negligible compared to the magnitude of QCD corrections that are discussed in this paper. It would also be easy to mitigate the effect of $b$ jets from the production subprocess with even a loose cut on the invariant mass of two $b$ jets, requiring it to select events which are sufficiently close to $m_H$. In this case, the impact of $b$ quarks from the WBF production subprocess on the cross section with $H \to b \bar{b}$ decay would be further reduced.
We take these observations as an indication that the impact of treating $b$ quarks from the production process as effectively unflavored should not qualitatively alter our findings.


\section{Conclusions}
\label{sec:conclusions}

Fiducial cross sections for Higgs boson production in weak boson fusion show bad perturbative convergence if scale uncertainties are used as a criterion for estimating higher-order QCD corrections \cite{Cacciari:2015jma,Cruz-Martinez:2018rod}. Considering the $H \to b\bar{b}$ decay at leading order, it was observed that additional $b$ jet selection criteria restrict the fiducial WBF region and improve the perturbative convergence~\cite{Asteriadis:2021gpd}.
However, an important effect, QCD radiation in the decay, was not included in that paper.

In the current paper, we extend that study by considering Higgs boson production in WBF followed by the decay of the Higgs boson into a pair of $b$ quarks, treating both production and decay at NNLO QCD. We found that WBF cross sections measured in the fiducial region akin to the one defined in Ref.~\cite{ATLAS:2017cen} are subject to large perturbative corrections, which reduce the LO cross section by about $\qty{40}{\percent}$. This large reduction appears as a result of a number of moderate effects that amplify each other. As the consequence, conclusions about the improved perturbative convergence reached in \cite{Asteriadis:2021gpd} no longer hold, at least for the set of cuts on $b$ jets that we have studied.

The transverse momentum cut on $b$ jets seems to play an important role in this because it causes QCD corrections to the $H \to b \bar{b}$ fiducial partial decay width to be large and different from the corrections to the inclusive one. Hence, if one hopes to use the $H \to b \bar{b}$ decay mode in WBF production to study the $b$ quark Yukawa coupling, it is crucial to control QCD corrections not only in the production subprocess but also to the decay. For better stability of perturbative computations, reducing the transverse momentum cut on $b$ jets appears to be useful, but whether this is possible given the fact that this will bring in additional backgrounds, requires a separate investigation.

Alternatively, one can try to resum perturbative corrections to the decay and/or match the result of this calculation to a parton shower. Since we observe a lack of perturbative stability in the decay of a color-singlet subject to fiducial constraints, it would be interesting to investigate this. As there is no color reconnection between production and decay processes, there should not be any conceptual obstacle to implementing such a calculation \cite{Bizon:2019tfo,Campbell:2021svd,Alioli:2020fzf}.

Finally, we note that a formal completion of the computation presented here would require modifying the existing NNLO computation for Higgs production in WBF to include the effect of massive $b$ quarks. While some parts of such a calculation are straightforward, the most nontrivial part will require a computation of NLO QCD correction to the case when a gluon splits into a $b\bar{b}$ pair and then one of the $b$ quarks from the splitting participates in the WBF process.
Such a calculation is definitely doable but, given the magnitude of the corrections found in this paper, it is highly unlikely to change our conclusions.


\section*{Acknowledgments}

We want to thank Wojciech Bizoń for a fruitful collaboration on the calculation of NNLO QCD corrections to $H \to b \bar{b}$ which we relied upon in this paper.
KA thanks Brookhaven National Laboratory, where a significant part of this research was conducted. RR has been partially supported by the Italian Ministry of Universities and
Research (MUR) through grant PRIN 2022BCXSW9.
The research of KM and IN is partially supported by the Deutsche Forschungsgemeinschaft (DFG, German Research Foundation) under grant 396021762~-~TRR~257.

\let\oldaddcontentsline\addcontentsline
\renewcommand{\addcontentsline}[3]{}
\bibliography{vbfhbb}
\let\addcontentsline\oldaddcontentsline

\end{document}